\documentclass[
twocolumn,
prl,showpacs,preprintnumbers,amsmath,amssymb]{revtex4}

\usepackage{graphicx}
\usepackage{amssymb}
\usepackage{amsmath}
\usepackage{dsfont}
\usepackage{bm}

\begin{document}

\title{Slow light, superadiabaticity and shape-preserving pulses}

\author{U. Leonhardt and L. C. D\'{a}vila Romero}

\affiliation{School of Physics and Astronomy, University of St
  Andrews, North Haugh, St Andrews, KY16 9SS, Scotland }     

\begin{abstract}
We analyze the nonlinear dynamics of 
atomic dark states in $\Lambda$ configuration
that interact with light at exact resonance.
We found a generalization of 
shape-preserving pulses
[R. Grobe, F. T. Hioe, and J. H. Eberly,
Phys. Rev. Lett. {\bf 73}, 3183 (1994)]
and show that the condition for
adiabaticity of the atomic dynamics is never violated,
as long as spontaneous emission is negligible.  
\end{abstract}


\maketitle

\indent
Slow light in atomic media \cite{Slow}
relies on the adiabatic following of atomic dark states 
\cite{AriBerg}.
The atoms are brought into such states by optical pumping 
\cite{Bernheim}, 
a dissipative process based on spontaneous emission.
Once they are in dark states, the atoms follow any changes 
in the applied light fields with remarkable ease \cite{Liu,Phillips}
and without further assistance by spontaneous emission,
much beyond the expectations of adiabatic theory \cite{Fleisch}.
In this note, we develop a brief argument as to why this is the case.

We consider the nonlinear dynamics
of light and atoms
in $\Lambda$ configuration with  
two co-propagating light beams at exact resonance,
see the figure.  
This case corresponds to slow and stopped light in
Bose-Einstein condensates \cite{Liu}.
Our analysis extends the nonadiabatic approach 
by Matsko {\it et al.} \cite{Matsko}
to a regime where the applied optical fields
can be of almost arbitrary strength,
whereas in their paper  \cite{Matsko} one
field dominates over the other.
As an important ingredient, we
construct a rather general approximate solution of the 
Maxwell-Schr\"odinger equations.
Our solution generalizes the 
shape-preserving pulses, or adiabatons, 
previously studied by
Grobe, Hioe, and Eberly \cite{Grobe}
and Fleischhauer and Manka \cite{FM},
to complex Rabi frequencies and complex time.
The solution also shows that 
in the case of zero detuning
not only slow-light solitons
\cite{LeoSlow,LambdaSolitons,Ruebe} 
represent stable structures
that can be slowed down, stopped
and accelerated depending on the total intensity.
In this case,
almost arbitrary optical polarization profiles 
can be stored and retrieved in atomic media,
as long as the atoms are not significantly excited. 
Using the continuation of our solution to complex time,
we show how the non-linear dynamics 
of atoms and light protects the atoms from
violating the adiabaticity condition \cite{LL3}.
One may call such behavior {\it superadiabaticity}.

\begin{figure}[t]
\begin{center}
\includegraphics[width=22.0pc]{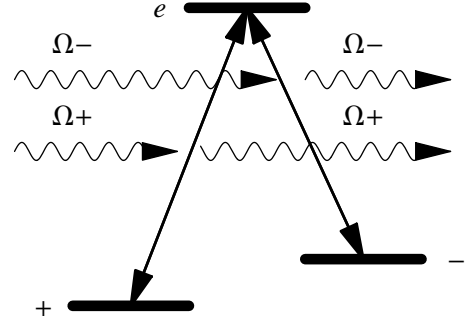}
 \caption{
\small
Two co-propagating light beams interact with a medium
composed of three-level atoms in $\Lambda$ 
configuration at exact resonance.
The electric field strengths of the light beams are
characterized by the Rabi frequencies 
\cite{Allen} $\Omega_\pm$.
The atomic levels are denoted by  
$|\pm\rangle$ and $|e\rangle$. 
} 
\end{center}
\end{figure}

Consider atoms in the $\Lambda$ configuration illustrated 
in the figure. 
We denote the lower atomic levels by 
$|\pm\rangle$ and the excited state by $|e\rangle$,
and assume that the atoms are in the pure states
$|\psi\rangle=\psi_+|+\rangle+\psi_-|-\rangle+\psi_e|e\rangle$.
The probability amplitudes may differ for atoms located
at different positions and they are time-dependent.
Suppose that the atoms are illuminated by two 
light beams  co-propagating in $z$ direction
with carrier frequencies that exactly match the
atomic transitions from the $|\pm\rangle$ to
the $|e\rangle$ level.
We describe the field strengths of the two polarizations
by the complex Rabi frequencies \cite{Allen} $\Omega_\pm$.
We assume that both the light and the atomic medium is uniform
in $x$ and $y$ direction. 
We introduce the retarded time $\tau=t-z/c$ and the 
scaled $z$ coordinate $\zeta=z/c$ that has the dimension of time.
In such coordinates, the light wave obeys the 
Maxwell equations 
in the slowly varying envelope approximation \cite{Allen}
%%%%%%
\begin{equation}
\label{eq:maxwell}
\partial_\zeta \Omega_\pm=
ig\,\bar{\psi}_\pm \psi_e
\,,\quad
\partial_\zeta \bar{\Omega}_\pm=
-ig\,\psi_\pm \bar{\psi}_e
\,,
\end{equation}
%%%%%%
where $g(\zeta)$ characterizes the atom-light interaction,
assumed to be equal for the $|\pm\rangle$ states,
that is proportional to the 
the atom-number density of the medium.
At this stage we neglect any atomic relaxation,
but we determine later the condition
when this is justified. 
Without relaxation the atoms obey the 
Schr\"odinger equation
%%%%%%
\begin{equation}
\label{eq:sch}
i\partial_\tau |\psi\rangle = H\,|\psi\rangle
\end{equation}
%%%%%%
with the Hamiltonian in the interaction picture 
%%%%%%
\begin{equation}
\label{eq:h}
H =
-\sum_\pm\bigg(
\frac{\bar{\Omega}_\pm}{2} |\pm\rangle \langle e| + 
\frac{\Omega_\pm}{2} |e\rangle \langle \pm|
\bigg)\,,
\end{equation}
%%%%%%
in the case of exact resonance of the atomic transition 
with the carrier frequencies of the light.

In the adiabatic dressed-state picture \cite{FM,LL3},
the atoms are assumed to be in instantaneous eigenstates
of the Hamiltonian (\ref{eq:h}),
in particular, in the atomic dark state \cite{AriBerg}
that depends on the applied optical fields.
When the fields vary the atoms follow
by remaining in the instantaneous eigenstate
while acquiring a phase that corresponds to 
the instantaneous eigenvalue integrated over time. 
However, when the atomic levels coincide the
atomic state is no longer well-defined and
adiabaticity is clearly violated,
because the atoms may easily go from one state to the other.
In practice, most level crossings are avoided in real time, 
but they may occur at certain points of complex time
for the analytically continued atomic dynamics.
In fact, these level crossings in the complex time plane
determine the Landau-Zener tunneling 
due to nonadiabaticity \cite{LL3}.
The exponential of the energy integrated 
from a complex level crossing to the real time axis
gives the tunneling probability amplitude  \cite{LL3}.

Consequemtly, in order to assess the adiabaticity
of the nonlinear dynamics
of light and atoms with
Hamiltonian  (\ref{eq:h})
we consider the analytic continuation of 
the Maxwell-Schr\"odinger equations 
(\ref{eq:maxwell}) and (\ref{eq:sch})
indicated by a subtlety of the notation:
the bar in $\bar{\Omega}_\pm$ and 
$\bar{\psi}_\pm$,  $\bar{\psi}_e$
denotes the analytic continuation of the
complex conjugate $\Omega^*_\pm$ and 
$\psi^*_\pm$,  $\psi^*_e$ from real to complex 
retarded time $\tau$ \cite{TrivialRemark1}.
Note that in general 
the analytic continuation of the complex conjugate
does not coincide with the complex conjugate
of the analytic continuation \cite{TrivialRemark1},
and therefore this distinction must be made.

The instantaneous eigenvalues of the Hamiltonian 
 (\ref{eq:h}) are
%%%%%%
\begin{equation}
\label{eq:eigen}
E_0 = 0 \,,\quad E_\pm = 
\pm\frac{1}{2}\sqrt{\bar{\Omega}_+\Omega_+  + 
\bar{\Omega}_-\Omega_-} \,.
\end{equation}
%%%%%%
The dark state is the eigenstate with the zero eigenvalue, 
a state that is orthogonal to the applied fields,
%%%%%%
\begin{equation}
\label{eq:dark}
\Omega_+\psi_+ + \Omega_-\psi_- = 0 \,.
\end{equation}
%%%%%%
Adiabaticity is violated in the vicinity of a level
crossing in complex time \cite{LL3},
unless the corresponding eigenstates are degenerate. 
In our case, level crossings occur when
%%%%%%
\begin{equation}
\label{eq:crossing}
\bar{\Omega}_+\Omega_+  + \bar{\Omega}_-\Omega_-
= 0 \,.
\end{equation}
%%%%%%
For real times, this implies that the fields vanish. 
In this case, the Hamiltonian (\ref{eq:h})
vanishes altogether and hence the eigenstates 
are degenerate.
No state mixing occurs.
On the other hand, the analytically continued  $\Omega_\pm$
and $\bar{\Omega}_\pm$ may satisfy the 
crossing condition for non-zero  $\Omega_\pm$ and
$\bar{\Omega}_\pm$, and in general they will 
\cite{TrivialRemark2}.
However, we show that the non-linear dynamics of
light and atoms prevents this from happening.

First, we generalize a previously known 
approximate solution of 
the atomic dynamics \cite{Leo} 
such that is is valid for complex time,
%%%%%%
\begin{eqnarray}
|\psi\rangle &=& N \left(|-\rangle - 
\frac{\Omega_-}{\Omega_+}|+\rangle +
\frac{2N_0^2}{\bar{\Omega}_+}
\partial_\tau \frac{\Omega_-}{\Omega_+}\,|e\rangle \right)
\,,\nonumber\\
N&=& N_0 \exp\left[
\frac{1}{2}\int N_0^2\left(
\frac{\Omega_-}{\Omega_+}\,\partial_\tau\,
\frac{\bar{\Omega}_-}{\bar{\Omega}_+} -
\frac{\bar{\Omega}_-}{\bar{\Omega}_+}\,\partial_\tau\,
\frac{\Omega_-}{\Omega_+}
\right) d\tau\right] \,,\nonumber\\
N_0&=&
\left(1 + \frac{\bar{\Omega}_-\Omega_-}
{\bar{\Omega}_+\Omega_+}\right) ^{-1/2} \,.
\label{eq:psi}
\end{eqnarray}
%%%%%%
We note that $N_0$ is invariant when $\Omega_\pm$ and
$\bar{\Omega}_\pm$ are exchanged, whereas
%%%%%%
\begin{equation}
\bar{N}= N_0 \exp\left[
\frac{1}{2}\int N_0^2\left(
\frac{\bar{\Omega}_-}{\bar{\Omega}_+}\,\partial_\tau\,
\frac{\Omega_-}{\Omega_+} - 
\frac{\Omega_-}{\Omega_+}\,\partial_\tau\,
\frac{\bar{\Omega}_-}{\bar{\Omega}_+}
\right) d\tau\right] \,.
\end{equation}
%%%%%%
The state vector corresponds to an evolving dark state
obeying the orthogonality condition (\ref{eq:dark}). 
The expressions (\ref{eq:psi}) exactly solve
the Schr\"odinger equation (\ref{eq:sch})
for the ground-state probability amplitudes $\psi_\pm$,
but not for the excited-state amplitude $\psi_e$.
This is justified, as long as the population of the upper level
is small in comparison with unity, a situation where
the lower states enslave the dynamics at the top level.

Given the solution (\ref{eq:psi}) of the atomic dynamics, 
we solve the Maxwell equations (\ref{eq:maxwell}) for the light
that turn into
%%%%%%
\begin{equation}
\label{eq:max2}
\partial_\zeta \Omega_+ = 
g\,\frac{2N_0^4}{\bar{\Omega}_+^2}\,
\bar{\Omega}_- \partial_\tau
\frac{\Omega_-}{\Omega_+}
\,,\quad
\partial_\zeta \Omega_- = 
-g\,\frac{2N_0^4}{\bar{\Omega}_+}\, 
\partial_\tau
\frac{\Omega_-}{\Omega_+}
\end{equation}
%%%%%%
and the corresponding equations for $\bar{\Omega}_\pm$.
We see that $\partial_\zeta(\bar{\Omega}_+\Omega_+  + 
\bar{\Omega}_-\Omega_-)$ vanishes, which implies
%%%%%%
\begin{equation}
\label{eq:omega2}
\bar{\Omega}_+\Omega_+  + \bar{\Omega}_-\Omega_- 
= \Omega^2(\tau) \,,
\end{equation}
%%%%%%
where $\Omega$ is real for real time.
We try the ansatz
%%%%%%
\begin{eqnarray}
\label{eq:ansatz}
\Omega_+ = \Omega\,e^{+i\phi_+}\cos\Theta
&,&\;
\Omega_- = \Omega\,e^{+i\phi_-}\sin\Theta
\,,\nonumber\\
\bar{\Omega}_+ = \Omega\,e^{-i\phi_+}\cos\Theta
&,&\;
\bar{\Omega}_- = \Omega\,e^{-i\phi_-}\sin\Theta
\,,
\end{eqnarray}
%%%%%%
which gives
%%%%%%
\begin{equation}
N_0=\cos\Theta \,.
\end{equation}
%%%%%%
One verifies that $\Theta$ and $\phi_\pm$ solve the
Maxwell equations (\ref{eq:max2}) when
%%%%%%
\begin{equation}
\label{eq:sol}
\Theta =\Theta(\xi) \,,\quad
\phi_+-\phi_- = \phi(\xi)
\end{equation}
%%%%%%
and
%%%%%%
\begin{equation}
\phi_++\phi_- = \int \cos(2\Theta)\,d\phi
\end{equation}
%%%%%%
with
%%%%%%
\begin{equation}
\label{eq:xi}
\xi = \zeta - \frac{1}{2g}\int\Omega^2 d\tau
\,.
\end{equation}
%%%%%%
Our solution generalizes a previous result 
by Grobe, Hioe, and Eberly \cite{Grobe}
to complex Rabi frequencies and complex time,
a prerequisite in assessing the adiabaticity 
of the system.
Equations (\ref{eq:omega2}) 
and (\ref{eq:ansatz}) imply that 
$\bar{\Omega}_+\Omega_+  + \bar{\Omega}_-\Omega_- $
vanishes at the zeros of $\Omega_\pm$ where
the Hamiltonian (\ref{eq:h}) vanishes.
Consequently, the levels are degenerate here 
and hence no adiabatic crossing occurs.

Our solution describes general shape-preserving pulses \cite{Grobe}.
The total envelope $\Omega(\tau)$ of the two matched light waves
propagates through the medium with the speed 
of light in vacuum, whereas the polarization structure
propagates with the characteristic speed of slow light \cite{Slow}
%%%%%%
\begin{equation}
v = c\,\frac{\Omega^2}{g}
\quad\mbox{for}\;g\gg \Omega^2
\,.
\end{equation}
%%%%%%
The total envelope plays the role of the control beam that 
sets the speed of the polarization pulse.
When $\Omega$ vanishes the structure is brought to a halt
and the polarization profile is stored in the atomic medium,
as long as it is not eroded by loss mechanisms.
Illuminating the medium sets the structure in motion again.
Therefore, in the case of zero detuning, arbitrary 
polarization profiles can be stored and retrieved,
not only slow-light solitons \cite{LeoSlow}, 
provided the profile does not significantly
excite the atoms ({\ref{eq:psi}).

To estimate the validity of our result,
we calculate $|\psi_e|^2$ for real retarded times.
We obtain from Eqs.\ (\ref{eq:psi}) and (\ref{eq:ansatz})
and the shape-invariant solution 
(\ref{eq:sol}) and (\ref{eq:xi})
%%%%%%
\begin{equation}
|\psi_e|^2 = \frac{\Omega^2}{4g^2}
\left[(2\partial_\xi\Theta)^2+
(\partial_\xi\phi)^2 \sin^2(2\Theta)\right]
\approx \frac{\Omega^2c^2}{g^2a^2}\,.
\end{equation}
%%%%%%
Here $a$ denotes the characteristic length scale
over which the polarization profile varies. 
Clearly, $|\psi_e|^2\ll 1$ when
$\Omega^2c^2 \ll g^2a^2$,
which limits the gradient of the polarization profile.
If this condition is not satisfied our
analysis is not applicable 
and the adiabaticity of the atomic dynamics may be violated.  
In practice, however, 
spontaneous emission from the excited state
usually poses a much more severe limitation.
Spontaneous emission amounts to losses 
and possibly the destruction of the Bose-Einstein condensate
in which the slow-light polarization structure is 
contained. To estimate the propagation length $l$ 
over which spontaneous emission is negligible, 
we follow the procedure of
Ref. \cite{LeoSlow} and obtain the loss rate
%%%%%%
\begin{equation}
\eta_L = 
\frac{32\pi}{n\lambda^3}\, 
\frac{l\lambda}{a^2}
\end{equation}
%%%%%%
where $\lambda$ denotes the optical wavelength.
For Bose-Einstein condensates \cite{Liu}
$n\lambda^3$ has been in the order of $1$ or larger.
Losses are thus negligible for pulses that are 
significantly longer than the geometric mean 
of wavelength and distance travelled.

{\it Summary.}
We derived analytic solutions 
for a general class of shape-preserving 
pulses interacting with
three-level atoms in $\Lambda$ configuration. 
Our theory indicates that 
the nonlinear dynamics of atoms and light 
protects the atoms from violating the 
adiabaticity condition,
as long as spontaneous emission 
is negligible. 
Slow light is superadiabatic.

We thank J. H. Eberly for the correspondence that inspired
this paper and we acknowledge the support of 
the Leverhulme Trust,
COVAQIAL,
and the Engineering and Physical Sciences Research Council.


\begin{thebibliography}{99}
\bibitem{TrivialRemark1}
To give a trivial example that shows that
the analytic continuation of the complex conjugate
$\bar{\Omega}_\pm$ 
does not coincide with the complex conjugate
$\Omega_\pm^*$,
suppose that $\Omega_+ = K \tau$
and $\Omega_- = K\tau_0$ with some real constants 
$K$ and $\tau_0$.
In this case,   $\bar{\Omega}_+=K \tau$,
whereas  $\Omega_+^*=K \tau^*$.
On the other hand,
$\bar{\Omega}_-=K\tau_0=\Omega_-^*$.
\end{thebibliography}
\end{document}